# Lithographic performance of ZEP520A and mr-PosEBR resists exposed by electron beam and extreme ultraviolet lithography

Roberto Fallica[a)], Dimitrios Kazazis

Paul Scherrer Institute, 5232 Villigen PSI, Switzerland

Robert Kirchner

Technische Universität Dresden, Institute of Semiconductors and Microsystems, 01062 Dresden, Germany

Anja Voigt

micro resist technology GmbH, Köpenicker Str. 325, 12555 Berlin, Germany

Iacopo Mochi, Helmut Schift, Yasin Ekinci

Paul Scherrer Institute, 5232 Villigen PSI, Switzerland

[a)] Electronic mail: roberto.fallica@psi.ch

Pattern transfer by deep anisotropic etch is a well-established technique for fabrication of nanoscale devices and structures. For this technique to be effective, the resist material plays a key role and must have high resolution, reasonable sensitivity and high etch selectivity against the conventional silicon substrate or underlayer film. In this work, the lithographic performance of two high etch resistance materials was evaluated: ZEP520A (Nippon Zeon Co.) and mr-PosEBR (micro resist technology GmbH). Both materials are positive tone, polymer-based and non-chemically amplified resists. Two exposure techniques were used: electron beam lithography (EBL) and extreme ultraviolet (EUV) lithography. These resists were originally designed for EBL patterning, where high quality patterning at sub-100 nm resolution was previously demonstrated. In the scope of this work, we also aim to validate their extendibility to EUV for high resolution and large area patterning. To this purpose, the same EBL process conditions were employed at EUV. The figures of merit, i.e.

dose to clear, dose to size, and resolution, were extracted and these results are discussed systematically. It was found that both materials are very fast at EUV (dose to clear lower than 12 mJ/cm$^2$) and are capable of resolving dense lines/space arrays with a resolution of 25 nm half-pitch. The quality of patterns was also very good and the sidewall roughness was below 6 nm. Interestingly, the general-purpose process used for EBL can be extended straightforwardly to EUV lithography with comparable high quality and yield. Our findings open new possibilities for lithographers who wish to devise novel fabrication schemes exploiting EUV for fabrication of nanostructures by deep etch pattern transfer.

## I. INTRODUCTION

Pattern transfer by deep anisotropic etching of a resist material into silicon substrate or underlayer has been an established technique to fabricate semiconductor structures and devices.[1] In this method, the resist material is first patterned, for example by electron beam or optical lithography, to change its solubility. Afterward, it is typically developed in positive tone (removing the exposed areas) and an anisotropic etch is performed to transfer the designed morphology into the underlying substrate. Owing to its high resolution and relative affordability in comparison to other techniques, electron beam lithography (EBL) has been widely adopted as resist patterning tool for etch transfer purposes in academic and small-scale research centers. In outlook, the scope of application of EBL is expected to expand even further as multi-beam systems become available and provide fast, high-resolution, large area patterning simultaneously.[2]

Although a wide range of inexpensive and sub-µm resolution polymer based-resists are available for positive tone processing, such as poly(methyl methacrylate) (PMMA), the low sensitivity and poor etch resistance make them unsuitable for pattern transfer purposes. Furthermore, the need for novel, smaller nanodevices and nanostructures keeps pushing the limits to the performance of resists and a great effort is devoted to the development of high-performance patternable materials from industry and academia.[3] To be successfully adopted, resist materials for nanoscale (< 100 nm) fabrication must demonstrate a combination of features: adequate resolution and low sidewall roughness, good sensitivity (to allow large-area patterning in a reasonable time) and sufficiently high mechanical properties, in order to withstand the damage of the chemical/physical etch.

Over the last two decades, ZEP520A prevailed as a highly successful resist both from a technological and a commercial point of view. It is a polymeric, non-chemically amplified resist specifically designed for EBL.[4] Extensive studies demonstrated its higher sensitivity and lower sidewall roughness in comparison to PMMA and novolak resin-based resists.[5] Moreover, its plasma etch selectivity to $SiO_2$ is about 2.9, which is significantly better than that of PMMA (0.4).[6,7] ZEP520A has become the material of choice for a variety of pattern transfer and sub-μm fabrication of memory devices,[8] transistors,[9] photonic crystals,[10] lithographic masks,[11,12] and for thermal reflow combined with grayscale lithography.[13]

Among the EBL resists for pattern transfer, mr-PosEBR is a recently released material based on a chlorine-containing copolymer with an aromatic group. mr-PosEBR has been recently demonstrated to have an etch selectivity of 2.3 to silicon (using $CF_4/SF_6$) and about as high sensitivity as its competitor, the ZEP520A. Notably, the resolution of mr-PosEBR reaches down to 35 nm so far[14] and can be used for grayscale lithography as well.[15] Both ZEP520A and mr-PosEBR are non-chemically amplified resists which makes them very stable with respect to shelf life and processing robustness.[16] Additionally, the post-exposure development of these materials is also easy to carry out and relatively straightforward with minimum requirements in terms of chemical facilities.

It is known that electron beam resists can also be exposed and patterned by extreme ultraviolet (EUV) light of 13.5 nm wavelength owing to the photoionization effect of such highly energetic (92 eV) photons.[17] Lithography at EUV is rising to great technological importance as it is expected to become the next generation platform for high volume manufacturing in the semiconductor industry.[18] The photoionization effect at EUV is fundamentally different to that triggered in EBL by electrons of tens of keV kinetic energy and it is the subject of extensive study;[19] however a comprehensive understanding is still lacking. Besides, very few works have experimentally investigated from the lithographic point of view the difference between EBL and EUV, and mostly to the purpose of calibrating the exposure results when EBL is used for fast resist screening of EUV resists.[20,21] On the other hand, while a broad choice of resists is available to electron-beam lithographers, there is a lack of EUV-specific resists for prototyping and small scale applications. Most of the research in EUV resist is proprietary and it is directed towards large scale (typically 12") and single-digit resolution with very thin ($\leq$ 25 nm) coating. There is therefore a substantial interest in assessing the patterning performance of non-chemically amplified resists at EUV provided that a good sensitivity (e.g. less than 20 mJ/cm$^2$) and good resolution (tens of nm) could be demonstrated. In this regard, the high etch selectivity of both ZEP520A and mr-PosEBR is the key feature to extend the scope of these materials to EUV lithography or to a mix-and-match combination of EUV and EBL.

In this work, we carried out a characterization of ZEP520A and mr-PosEBR sensitivity to both electron beam and EUV radiation, under the same processing conditions. The objective is to demonstrate the feasibility of these materials at EUV, specifically with regard to the exposure sensitivity and patterning quality. The comparison of sensitivity also let us draw some conclusions about the effective energy required to pattern a given material using electrons as opposed to photons. Moreover, high resolution features in dense arrays of lines/space (l/s) were produced and their morphology was evaluated by scanning electron microscopy. The relevant figures of merit for lithographic quality were compared to the state-of-the-art in EBL and EUV. It is of interest to determine whether the well-established process conditions of EBL could be extended to a radically different platform as EUV. It is expected that these results will open new possibilities in micro- and nano-fabrication with ZEP520A and mr-PosEBR and provide useful insight to manufacturers in the field who wish to explore novel applications.

## II. EXPERIMENTAL DETAILS

Two resist materials, ZEP520A and mr-PosEBR have been investigated in this work. The former is produced by Nippon Zeon Co. and consists of a 1:1 ratio copolymer of chloromethacrylate and methylstyrene with molecular weight of 57 000 g/mol, and anisole as casting solvent.[4] The latter, developed by *micro resist technology* GmbH, consists of an organic copolymer of an acrylic, chlorine-containing monomer with an acrylic monomer bearing aromatic side-group (total molecular weight 134 200 g/mol).[14] An additional resist material was also used as reference for the comparison of the lithographic sensitivity. To this purpose, PMMA of molecular weight 950 000 g/mol and concentration 1% w/w, cast in ethyl lactate, was used. Notably, all of these materials are sensitive to exposure by both electrons and EUV photons over a broad range of energies.

An EBL tool (Vistec EBPG 5000 Plus) with 100 keV acceleration voltage, beam current of 250 pA, beam aperture of 200 μm, was used to study the resist response to electron beam exposure. The sensitivity to electron beam was determined using the conventional method of contrast curves. To this end, large areas (100 × 100 μm$^2$) of PMMA, ZEP520A and mr-PosEBR were exposed to increasing doses of electron beam from 10 to 1000 μC/cm$^2$. After development, the remaining resist thickness was measured by stylus profilometry (relative to the thickness of the unexposed areas). The experimental thickness data was fitted to a sigmoidal dose response curve using a method described elsewhere.[22] The dose to clear 50%, figure of merit of the resist sensitivity, was thus determined from the sigmoidal curve as the dose at which the remaining thickness was 50% of the initial film thickness. The lithographic contrast, γ, was also extracted from the data as the slope of the derivative of the thickness vs. log$_{10}$(dose), evaluated at the dose to clear. The EBL patterning performance of ZEP520A and mr-PosEBR was evaluated by exposure of dense l/s arrays over a 4 × 4 μm$^2$ area. The nominal half-pitch of lines ranged from 50 to 22 nm and the dose varied from 125 to 1000 μC/cm$^2$.

Optical lithography at EUV wavelength was carried out at the XIL-II beamline of the Swiss Light Source synchrotron.[23] The EUV sensitivity of PMMA, ZEP520A and mr-PosEBR was determined, in a similar fashion, by the contrast curves method. In this case, flood exposures of 0.5 × 0.5 mm$^2$ area and dose ranging from 0.1 to 100 mJ/cm$^2$ were carried out. The remaining thickness of resist (after development) was measured as a function of dose, and relative to the thickness of unexposed areas. The lithographic contrast, γ, was also extracted from the data as the slope of the derivative of the thickness vs. $\log_{10}$(dose), evaluated at the dose to clear. For patterning tests, the EUV interference lithography (EUV-IL) scheme was employed. In EUV-IL, masks featuring transmission diffraction gratings produce two-beam interference patterns on the surface of the sample. ZEP520A and mr-PosEBR were exposed in dense l/s arrays of several tens of square microns of area, and half-pitch of 50, 25 and 22 nm with dose ranging from 10 to 150 mJ/cm$^2$. Further details on this methodology have been described elsewhere.[24] The main advantages of the EUV-IL technique are the high resolution (down to 6 nm half-pitch)[25], the large area patterning, and the short exposure time.

The quality and morphology of the l/s patterns obtained by EBL and EUV-IL was inspected by use of top-down scanning electron microscopy (SEM). The feature size was quantitatively extracted by metrological analysis of the SEM images using SUMMIT software suite (Lithometrix, USA). Consistently with established terminology, the feature size (linewidth) is indicated as critical dimension (CD); the stochastic variation of the sidewall is indicated as line edge roughness (LER). Besides, the quantitative metrology also provides the dose to size, which is the relevant figure of merit and represents the exposure dose at which the ratio between line width and trench width is 1:1.

For the sake of this comparison, the processing conditions were kept the same throughout all lithographic tools: thermal treatments, development time, and development conditions. Samples were prepared by spin coating the resists on bulk silicon substrates (550 μm thick, 4" wafers) to obtain films of 40 nm thickness. To this purpose, resist concentration and spin speed were tuned accordingly. This choice of thickness was dictated by the inability of patterning thick films by EUV light due to its short penetration depth in matter.[26] A post application bake of 140 °C, for 120 s, was used for ZEP520A and mr-PosEBR, while PMMA was baked at 180 °C, for 300 s. After exposure, ZEP520A and mr-PosEBR were developed in electronic-grade amyl acetate (≥ 99 %, Sigma-Aldrich) for 20 s, and subsequently rinsed in isopropyl alcohol (IPA) for 20 s. We note that amyl acetate is the manufacturer's designated developer for ZEP520A.[4] The PMMA samples were developed in a mixture of methyl isobutyl ketone and isopropyl alcohol of ratio 1:3 (MIBK:IPA) for 30 s, and rinsed for 30 s in IPA. All developments were carried out at room temperature. After development and rinse, the samples were blow-dried with ultrapure nitrogen to remove any residual rinsing liquid. The aforementioned processing conditions are summarized in Table I.

TABLE I. Process conditions of PMMA, ZEP520A and mr-PosEBR resist materials explored in this work: post application bake (PAB) temperature and time, developer solvent and development duration, rinse type and duration. Identical conditions were used to prepare samples for both EBL and EUV-IL exposures.

|  | PAB [°C] | PAB [s] | Developer | Dev. time [s] | Rinse | Rinse time [s] |
|---|---|---|---|---|---|---|
| PMMA | 180 | 300 | 1:3 MIBK:IPA | 30 | IPA | 30 |
| ZEP520A | 140 | 120 | Amyl acetate | 20 | IPA | 20 |
| mr-PosEBR | 140 | 120 | Amyl acetate | 20 | IPA | 20 |

## III. RESULTS AND DISCUSSION

### A. *Contrast curves and sensitivity*

In positive tone processing, the exposure to photons or electrons results in an increase of the solubility of the polymer-based resist owing to a chain scission mechanism. Indeed, the exposure dose needs to be calibrated to be high enough to trigger the solubility switch, but not disproportionate to prevent the re-polymerization of the material. After development, the remaining thickness, in the exposed areas, decreases owing to the (partial or total) dissolution of the resist film. The remaining thickness of resist therefore decreases at increasing dose and the experimental relation between these two quantities provides the lithographic sensitivity and allows for a direct comparison between resists and exposure processes. From the contrast curves, the dose to clear (which is the figure of merit that describes the sensitivity of a given material when exposed to photons and electrons) is then extracted. Fig. 1 (a) and (b) show the experimental contrast curves of the resist materials under study, exposed to light at EUV wavelength and to 100 keV electrons, respectively. In the plot, the thickness was normalized to the initial resist thickness, which was 40 nm for all samples. The dose to clear at 50% thickness and the lithographic contrast γ, which were extracted from the best fitting dose response curve, are summarized in Table II.

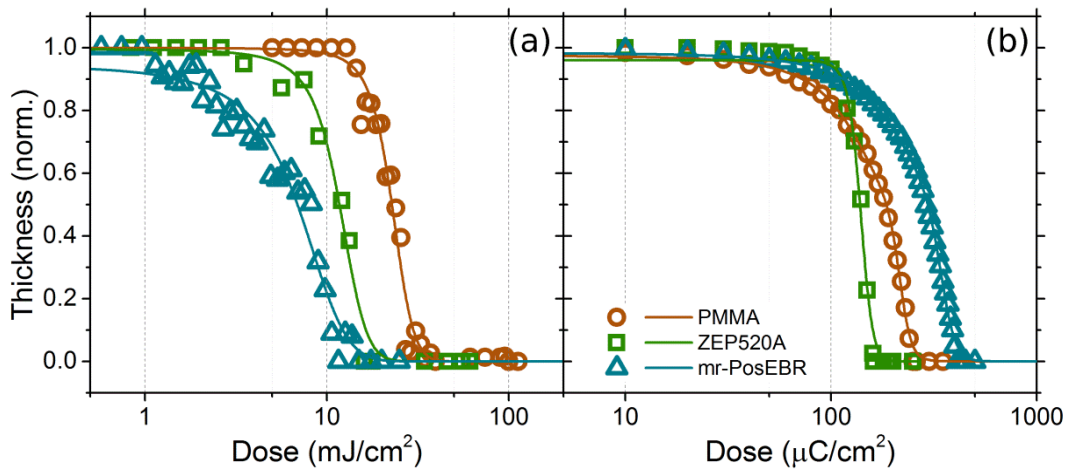

FIG. 1. (Color online) Experimental contrast curves (symbols) and best fitting based on a dose response curve (lines) of resist materials PMMA (brown circles), ZEP520A (green squares) and mr-PosEBR (cyan triangles) exposed to extreme ultraviolet wavelength photons (a) and 100 keV electron beam (b). The remaining resist

thickness, measured by stylus profilometry, was normalized to the initial thickness which was 40 nm for all samples.

At EUV, both ZEP520A and mr-PosEBR demonstrated good sensitivity: the dose to clear was 11.7 and 6.9 mJ/cm$^2$ respectively, outperforming PMMA (at 22.8 mJ/cm$^2$). This result shows that non-chemically amplified polymeric materials are capable of achieving a remarkable sensitivity which makes them promising candidates for use at EUV wavelength. The contrast curves obtained by electron beam exposure (Fig. 1(b)) showed a different behavior among the three materials. Here, the mr-PosEBR exhibited the lowest sensitivity, its dose to clear being more than twice that of ZEP520A. The ZEP520A was faster than PMMA both at EBL and EUV, although its superiority was less marked at EBL (a factor 1.3 faster) than it was at EUV (almost twice as fast). The high sensitivity of ZEP520A with respect to PMMA is consistent with previous studies where amyl acetate was used as developer[5] and at 75 keV acceleration voltage.[20] Other authors reported that a much lower dose of ≈ 20 μC/cm$^2$ was needed when using a 10 keV acceleration voltage,[31] as one would expect.[27]

As for the mr-PosEBR, since the samples were coated at the same initial thickness and processed under the same development conditions, the different behavior must be ascribed to differences in the exposure mechanism. In general, the amount of deposited dose needed to clear the resist film depends on the relative efficiency with which each photon or electron contributes to increasing the polymer solubility. A comprehensive understanding of electron-matter interaction is still lacking for low energy electrons, i.e., specifically for EUV photoelectrons (whereas a good model exists from classical approximations, for example in the Bethe theory[28]) and is beyond the scope of this work. As far as the lithographic sensitivity is concerned, a previous study proposed that the dose ratio between EUV and EBL is constant and can be estimated from simulations[21], although conclusive experimental evidence is still lacking. In the broad variety of resist materials, the sensitivity ratio is not always constant (despite using the same operational conditions) which might suggest that this ratio is a material-specific property. As for mr-PosEBR, it could be speculated that a higher density of the coated film might determine a higher absorption of EUV radiation and thus a higher solubility switch upon exposure, whereas the film density would be irrelevant when 100 keV electrons are used. Furthermore, there are at least two ways in which the chemical composition is expected to play a role in the solubility switch. First, the amount and configuration of chlorine within the polymer has been found to facilitate its scission, as previous studies found.[29] Second, the cleavage mechanism in acrylate polymers, such as mr-PosEBR, proceeds by formation of free radicals (Norrish Type I cleavage) opposite to what happens in methacrylate, where abstraction of a gamma-hydrogen (Norrish Type II) prevails.[30] All these effects could, in varying degree, concur with our experimental observation that EUV radiation is significantly more effective in clearing mr-PosEBR than it does with the two other methacrylate-based resists, at the same nominal incident dose.

As for the lithographic contrast, higher $\gamma$ indicates a broader gap in the development rate between the exposed areas vs. the unexposed areas. All the materials investigated in this work yielded a higher contrast using the EBL than they did when EUV light was used. ZEP520A and mr-PosEBR performed definitely better at EBL where their $\gamma$ was ≈ 2.3 times higher than it was at EUV. PMMA, instead, had only about 12% more contrast in the electron beam exposure. This finding suggests that the exposure mechanism of the two commercial products is optimized to enhance the solubility switch of the resist upon exposure to the highly energetic electron beam.

High γ at EBL is also a specific advantage for applications where the patterning of deep, high aspect ratio trenches is required.

TABLE II. Dose to clear at 50% thickness and lithographic contrast γ of PMMA, ZEP520A, and mr-PosEBR resist exposed by EUV and EBL. Values extracted from the contrast curves of Fig. 1.

|  | *Dose to clear 50 %* |  | *γ* |  |
| --- | --- | --- | --- | --- |
|  | EUV [mJ/cm$^2$] | EBL [μC/cm$^2$] | EUV | EBL |
| PMMA | 22.8 | 186 | 4.1 | 4.6 |
| ZEP520A | 11.7 | 138 | 3.3 | 8.0 |
| mr-PosEBR | 6.9 | 295 | 1.8 | 4.0 |

## *B. Patterning performance*

A great variety of processing conditions have been explored by EBL and reported for PMMA and ZEP520A,[5,13,31] and for mr-PosEBR.[14] The performance of PMMA is well known at EUV too.[32,33] In the scope of this work, we focus on the comparison of dense l/s array patterning by EBL and EUV-IL under the same processing conditions. Furthermore, it is of interest to demonstrate the printability of these resists at reasonable doses in the EUV regime while maintaining the general-purpose development conditions specified by the manufacturer.

The morphology of all samples was analyzed by top-down scanning electron microscopy and is shown, for selected doses, in the SEM images of Fig. 2. Some important conclusions can be drawn from the SEM inspection. Regardless of exposure tool, high quality patterns were achieved for both resists at half-pitch 50 nm. The l/s patterns were still acceptable at half-pitch 25 nm. However, only ZEP520A was capable of resolving features at 22 nm half-pitch, at which resolution the mr-PosEBR performed poorly. Regardless of the exposure tool, the failure mechanism of both resists was identified as bridging between adjacent lines, in other words the inability to completely clear the exposed trenches while preserving the integrity of the unexposed lines. No other failure mechanisms and in particular no pattern collapse was detected, most probably due to the low aspect ratio of the lines (40 nm height / 22 nm width). Although both resists can easily achieve high aspect ratio structures by EBL,[4,14] the short optical absorption depth of EUV wavelength does not allow the patterning of deep trenches.

|  | ZEP520A | | mr-PosEBR | |
|---|---|---|---|---|
|  | EUV-IL | EBL | EUV-IL | EBL |
| hp50 | 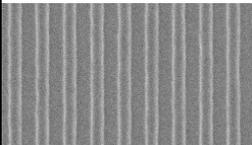 78 mJ/cm$^2$ | 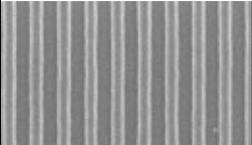 250 μC/cm$^2$ | 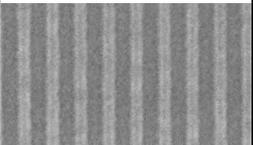 85 mJ/cm$^2$ | 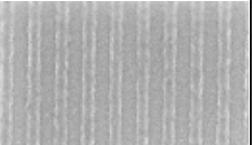 410 μC/cm$^2$ |
| hp25 | 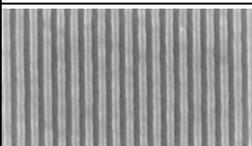 78 mJ/cm$^2$ | 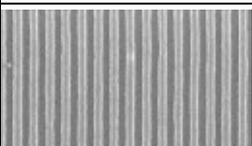 250 μC/cm$^2$ | 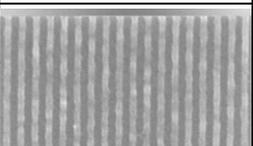 85 mJ/cm$^2$ | 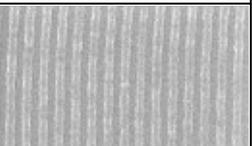 330 μC/cm$^2$ |
| hp22 | 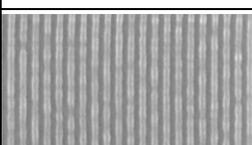 78 mJ/cm$^2$ | 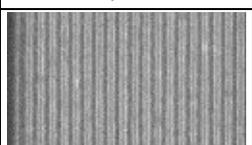 290 μC/cm$^2$ | 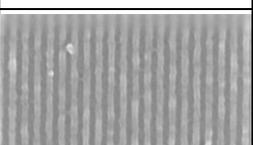 94 mJ/cm$^2$ | 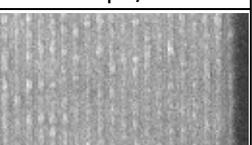 330 μC/cm$^2$ |

FIG. 2. Selected scanning electron micrographs of l/s patterns of ZEP520A and mr-PosEBR at half-pitches of 50 nm, 25 nm and 22 nm. The exposure dose of each sample is specified below each picture. Image areas are $1.1 \times 0.6$ μm$^2$.

From the quantitative analysis of SEM images we extracted the relevant figures of merit: CD vs. dose, and dose to size. CD vs. dose plots are shown in Fig. 3

(a) and (b) for EUV-IL and EBL, respectively. Limited data points are available at half-pitch 22 and 25 nm which, as observed from SEM inspection, are the patterning limits of ZEP520A and mr-PosEBR, respectively. The dose to size was extracted by linear interpolation of the experimental CD vs. dose data at values of CD equal to the half-pitch. In our samples of half-pitch 22 nm and 25 nm, the experimentally measured CD (linewidth) was larger than the half-pitch, owing to the limitations of the resists in resolving smaller features. Nonetheless, the extracted dose to size still holds valid and it is summarized in Table III. It can be noticed that the dose required to achieve a 1:1 line/space width ratio increases at smaller pitches, consistently with previous studies.[24] The smallest feature size we achieved in ZEP520A was 23.9 nm (EUV-IL) and 22.9 nm (EBL). We note that high resolution is more challenging to achieve in dense arrays than it is in isolated features. Moreover, we employed general-purpose process conditions which are not optimized for high resolution (for instance, because the development was done at room temperature). Other studies have shown that improved processing can yield trench width as small as 19 nm[5]; further improvement can be also achieved using cold development, with which 14 nm wide lines have been resolved.[34] As for the mr-PosEBR, other authors reported the successful patterning of smooth arrays of 29 nm wide lines by EBL in a film of thickness 320 nm (i.e., a very high aspect ratio of 11).[14] Our findings extend the process window of mr-PosEBR towards the high resolution and demonstrate that feature size as small as 24.4 nm (EUV-IL) and 25.4 nm (EBL) can be obtained, although the l/s were not as smooth as in ZEP520A. This result is promising in view of the refinement of development conditions towards higher resolution. Overall, both ZEP520A and mr-PosEBR showed excellent compatibility and good yield with both tools and demonstrated how the process conditions – optimized for EBL – can be extended to EUV-IL seamlessly.

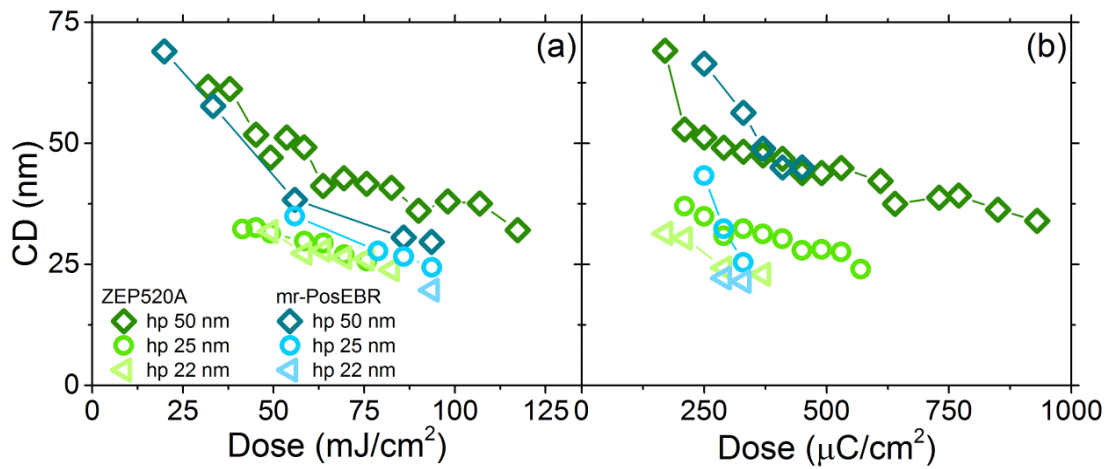

FIG. 3. (Color online) Critical dimension (CD) vs. dose plots for l/s patterns of ZEP520A (green symbols) and mr-PosEBR (cyan symbols) resists of half-pitch 50 nm (diamonds), 25 nm (circles) and 22 nm (triangles) obtained by EUV-IL (a) and EBL (b).

Finally, it is worthwhile to comment on the relation between the dose to size reported here and the dose to clear reported in the previous section. The sensitivity gap between ZEP520A and mr-PosEBR (that was found using the contrast curves) became negligible in EUV-IL patterning, where the dose is almost the same. The EBL pattern exposure of mr-PosEBR was about 1.2 times slower than it was for ZEP520A – consistently with the dose to clear trend discussed earlier.

TABLE III. Summary of the extracted dose to size, per half-pitch (hp) in nm, of ZEP520A and mr-PosEBR exposed by EUV-IL and EBL.

| | *EUV-IL [mJ/cm$^2$]* | | | *EBL [µC/cm$^2$]* | | |
|---|---|---|---|---|---|---|
| | *hp50* | *hp25* | *hp22* | *hp50* | *hp25* | *hp22* |
| ZEP520A | 55.5 | 79.2 | 89.6 | 270 | 410 | 450 |
| mr-PosEBR | 52.5 | 88.7 | 95 | 365 | 425 | 490 |

The sidewall roughness of a resist, expressed by the LER figure of merit, is relevant for pattern transfer applications as it directly affects the quality of the morphology etched into the underlying layers or substrate. It represents the statistic deviation (1 σ) of the sidewall contours that a real line exhibits as compared to that of an ideally straight line. LER is a critical parameter in high volume manufacturing as well: the International Roadmap for Semiconductors constrains its maximum allowable value on the basis of the technological node. The experimental LER of our resists, obtained by metrology of the l/s arrays, is shown in Fig. 4. Some conclusions can be drawn from our experimental observations. In the first place, the ZEP520A produced consistently smoother l/s arrays than the mr-PosEBR did. In particular, the ZEP520A achieved a remarkable ≈ 1.5 nm and ≈ 2 nm minimum LER (at the dose to size) using EUV-IL and at EBL respectively. The minimum LER that we could obtain with mr-PosEBR was limited to a narrower range of doses and was, at best, ≈ 4 nm (EUV-IL) and ≈ 6 nm (EBL). The roughness is also to be ascribed to the polymer size and thus to its radius of gyration, as was found in a previous study, where it was found that PMMA of large molecular weight (950 000) also resulted in > 6 nm LER.[32] Finally, the exposure tool also plays a role in the formation of the resist image after development. The normalized image log-slope (NILS) is the figure of merit for the quality of the aerial image. The higher the NILS, the sharper the slope of the aerial image intensity across the boundary between exposed and unexposed areas is. The consequence of a good aerial image is substantial especially when patterning dense arrays of alternating lines (unexposed) and spaces (exposed) arrays. For the EUV-IL tool, NILS is constant at $\pi$ regardless of pitch.[22] The intensity profile of a single, well collimated electron beam can have a higher NILS than $\pi$. However, the patterning by EBL occurs by stepping the beam at 4 nm-apart intervals (as in our case). The resulting aerial image of EBL is effectively the sum of a number of gaussians across the line width (from 4 to 12 depending on the target CD). It is easily demonstrated that the NILS of one gaussian is approximately the same as that of a sinusoid having equal FWHF; but as more and more gaussians are summed, their NILS becomes worse. As a result, the aerial image formed by a single sinusoid (EUV-IL) of outperforms the EBL obtained by beam stepping and contributes to patterning lines with low LER. A derivation of this theory also clarifies the reason why the LER of l/s arrays has its minimum when exposure dose is equal to the dose to size, that is, when the NILS is highest. Other factors, besides the NILS, are sources of LER. Among these, the different shot noise statistics of EUV photons and 100 keV electrons have been investigated. Notably, it was also found that EBL produces as much as 2.5-fold larger LER than EUV does, despite having the same nominal aerial image.[35] All these considerations explain why both resists studied in this work consistently yielded a lower LER at EUV-IL than they did at EBL.

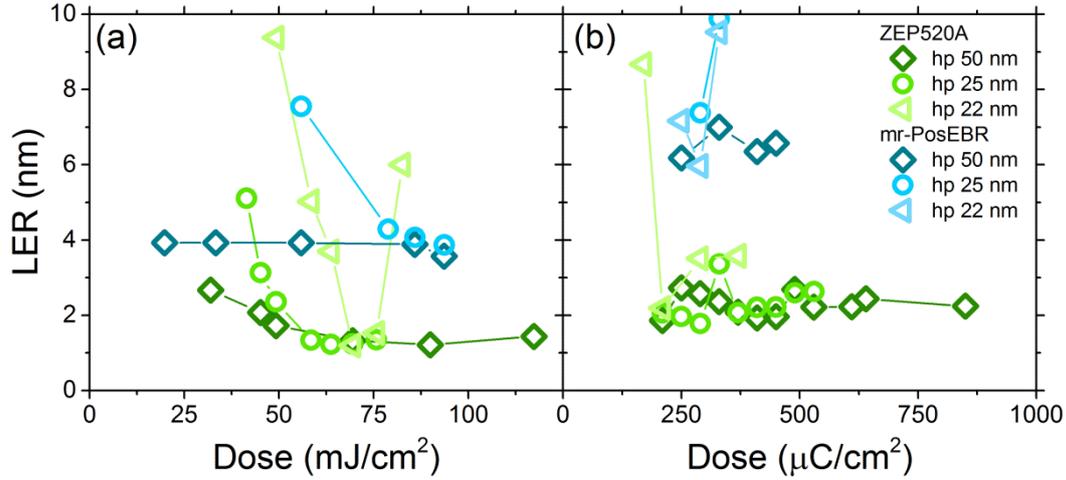

FIG. 4. (Color online) Line edge roughness (LWR) vs. dose plots for l/s patterns of ZEP520A (green symbols) and mr-PosEBR (cyan symbols) resists of half-pitch 50 nm (diamonds), 25 nm (circles) and 22 nm (triangles) obtained by EUV-IL (a) and EBL (b).

## IV. SUMMARY AND CONCLUSIONS

We have studied ZEP520A and mr-PosEBR, two polymer-based resists with high etch resistance originally designed for electron beam lithography. Interestingly, these materials showed excellent sensitivity to EUV radiation, where they both outperform PMMA. The high sensitivity brings the important advantage of reduced exposure time in those applications where large area patterning is required. Moreover, we have tested the patterning quality using both EBL and EUV-IL, where these resists yielded excellent dense arrays of 50 nm hp resolution, and down to 25 nm hp with acceptable LER. This finding demonstrates that a general-purpose process used for EBL can be extended dependably to the EUV-IL achieving a comparably high resolution. These results, along with other features as affordability, wide availability and improved etch resistance open new opportunities for nano- and micro-fabrication with ZEP520A and mr-PosEBR. For instance, EUV-IL could be combined with EBL in a mix-and-match process for fast and large area patterning of arbitrary and periodic high resolution structures at once. The chemical formulation of these materials, specifically the absence of photoacid chemistry, guarantees long shelf life, long term stability and robustness against process fluctuations.[16] In summary, it is expected that lithographers will be able to implement a broad range of applications which will benefit from the fast, cost-effective, high resolution patterning for prototyping and device manufacturing.

## ACKNOWLEDGMENTS

Michaela Vockenhuber and Vitaliy Guzenko (PSI) are kindly acknowledged for technical support and fruitful discussions. This project has received funding from the EU-H2020 Research and Innovation program under Grant Agreement No. 654360 NFFA-Europe.